\documentstyle[12pt,epsf,epsfig,here]{article}
\def\beq{\begin{equation}}
\def\eeq{\end{equation}}
\def\bea{\begin{eqnarray}}
\def\eea{\end{eqnarray}}
\def\bq{\begin{quote}}
\def\eq{\end{quote}}

\def\CMP{{\it Commun.Math.Phys.} }

\def\SNC{{\it Suppl. Nuovo Cimento} }

\def\NC{{\it Nuovo Cimento} }
\def\NP{{\it Nucl.Phys.} }
\def\PL{{\it Phys.Lett.} }
\def\PR{{\it Phys.Rev.} }
\def\PRL{{\it Phys.Rev.Lett.} }

\def\gappeq{\mathrel{\rlap {\raise.5ex\hbox{$>$}}
{\lower.5ex\hbox{$\sim$}}}}

\def\lappeq{\mathrel{\rlap{\raise.5ex\hbox{$<$}}
{\lower.5ex\hbox{$\sim$}}}}

\begin{document}
\pagestyle{empty}
\begin{flushright}
CERN-TH/99-110\\
LAPTH-737/99
\end{flushright}
\vspace*{5mm}
\begin{center}
{\bf THE RIGOROUS ANALYTICITY-UNITARITY PROGRAM}\\
{\bf AND ITS SUCCESSES}
\\
\vspace*{1cm}
{\bf Andr\'e MARTIN} \\
\vspace{0.3cm}
Theoretical Physics Division, CERN \\
CH - 1211 Geneva 23 \\
and
\\
LAPP\footnote{URA 1436. Associ\'e \`a l'Universit\'e de Savoie.} - F 74941
ANNECY LE VIEUX\\ e-mail: {\tt martina@mail.cern.ch}\\
\vspace*{2cm}
{\bf ABSTRACT} \\
\end{center}

We show how the combination of analyticity properties derived from local field
theory and the unitarity condition (in particular positivity) leads to
non-trivial physical results, including the proof of the ``Froissart bound"
from first principles and the existence of absolute bounds on the pion-pion
scattering amplitude.
\vspace*{1cm}

\begin{center}
{\it Talk given at the}\\
{\it Ringberg Symposium on Quantum Field Theory}\\
{\it in honour of Wolfhart Zimmermann}\\
{\it Ringberg Castle, Tegernsee, Germany}\\
{\it June 1998}
\end{center}

\vspace*{1cm}
\begin{flushleft}
CERN-TH/99-110\\
LAPTH-737/99\\
April 1999
\end{flushleft}
\vfill\eject
\pagestyle{plain}
\setcounter{page}{1}

I would like to begin by wishing a very happy birthday to Wolfhart Zimmermann.
I have chosen a topic which is close to the interests of Wolfhart and as you
 will see soon, in which Wolfhart has made a crucial contribution which makes
all the work made in the ``pre-quark" era still valid now.

My task would have been much easier if the scheduled first speaker of this
conference, Harry Lehmann, had been present. Unfortunately he was ill, and,
at the time of writing this talk, we know that he left us. As we shall see all
through what follows, the contributions of Harry Lehmann to that domain are
many and all of them are fundamental.

In 1954, Gell-Mann, Goldberger and Thirring \cite{aaa} proved that dispersion
relation, previously developed in optics could be established for Compton
Scattering: $\gamma P \rightarrow \gamma P$, from the existence of local 
fields satisfying the causality property
$$
[A(x), A(y)] = 0~~{\rm for}~~ (x-y)^2 < 0~,
$$
i.e., spacelike. This made it possible to express the real part of the forward
scattering amplitude as an integral over the imaginary part of the forward
scattering amplitude, i.e., by the ``optical theorem", an integral over the
total cross-section for Compton Scattering. At  the same time a general
formulation of quantum field theory incorporating causality giving in
particular general expression for scattering amplitude was developed by
Lehmann, Zimmermann and Symanzik (LSZ) in their pioneering paper (in german!)
in Nuovo Cimento \cite{bb}.

On this basis, dispersion relations were ``proposed" for massive particles in
the work of Goldberger on the pion-nucleon scattering amplitude \cite{cc}.
Soon, his ``heuristic proof" was turned into a real proof by various authors
using the LSZ formalism \cite{dd}. One of these proofs is due again to Harry
Lehmann!

	Before going on, I would like to explain that if these results, even after the
discovery that protons and pions are not elementary but made of quarks, are
still valid, it is thanks to a fundamental contribution of Wolfart Zimmermann
entitled ``On the bound state problem in quantum field theory" \cite{ee}, in
which it is proved that to a bound state we can associate a local operator.
This constitutes an excellent answer to sceptics like Volodia Gribov \cite{ff}
or Klaus Hepp \cite{ggg} (qui br\^ule ce qu'il a ador\'e!). 

Now I believe that it is necessary to give some technical details, even if most
of you know about it.

In 3+1 dimensions (3 space, 1 time) the scattering amplitude depends on two
variables energy and angle. For a reaction $A + B \rightarrow A + B$
\beq
E_{c.m.} = \sqrt{M^2_A+k^2} + \sqrt{M^2_B+k^2}~,
\label{one}
\eeq
$k$ being the centre-of-mass momentum. The angle is designated by $\theta$.
There are alternative variables:
\beq
s = (E_{CM})^2~,~~~t = 2k^2 (\cos\theta -1)
\label{two}
\eeq(Notice that physical $t$ is NEGATIVE).

We shall need later an auxiliary variable $u$, defined by '\beq
s + t + u = 2 M_A^2 + 2M^2_B
\label{three}
\eeq

The \underline{Scattering amplitude} (scalar case) can be written as a partial
wave expansion, the convergence of which will be justified in a moment:
\beq
F (s,\cos\theta ) = {\sqrt{s}\over k} \sum (2\ell +1) f_\ell(s) P_\ell
(\cos\theta)
\label{four}
\eeq
$f_\ell(s)$ is a partial wave amplitude.

The \underline{Absorptive part}, which coincides for $\cos\theta$ real 
(i.e., physical) with the imaginary part of $F$, is defined as
\beq
A_s (s,\cos\theta ) = {\sqrt{s}\over k} \sum (2\ell + 1) ~{\rm Im}~ f_\ell (s)
(\cos\theta)
\label{five}
\eeq

The \underline{Unitarity condition}, implies, with the normalization we have
chosen
\beq
{\rm Im}~ f_\ell (s) \geq  \vert f_\ell (s) \vert^2
\label{six}
\eeq
which has, as a consequence
\beq
{\rm Im}~ f_\ell (s) > 0~,~~~ \vert f_\ell  \vert < 1~.
\label{seven}
\eeq
The differential cross-section is given by
$$
{d\sigma\over d\Omega} = {1\over s}~~\vert F\vert^2~,
$$
and the total cross-section is given by the ``optical theorem"
\beq
\sigma_{total} = {4\pi\over k\sqrt{s}}~~A_s(s,\cos\theta = 1)~.
\label{eight}
\eeq

With these definitions, a dispersion relation can be written as:
\beq
F(s,t,u) = {1\over\pi} \int{A_s(s^\prime,t)ds^\prime\over s^\prime -s} +
{1\over\pi}
\int {A_u(u^\prime,t)du^\prime\over u^\prime -u}
\label{nine}
\eeq
with possible subractions, i.e., for instance the replacement of $1/(s^\prime
- s)$ by
$s^N /s^{\prime N}(s^\prime - s)$
and the addition of a polynomial in $s$, with coefficients depending on $t$. 

The scattering amplitude in the $s$ channel $A+B\rightarrow A+B$ is the
boundary value of $F$ for $s + i\epsilon$, $\epsilon > 0  \rightarrow 0$, $s >
(M_A+M_B)^2$. In the same way the amplitude for $A+\bar B\rightarrow A+\bar B$, 
$\bar B$ being the antiparticle of $B$ is given by the boundary value of $F$
for $u+i\epsilon , ~~\epsilon\rightarrow 0 ~~u > (M_A+M_B)^2$. Here we
understand the need for the auxiliary variable $u$.

The dispersion relation implies that, for fixed $t$ the scattering amplitude
can be continued in the $s$ complex plane with two cuts. The scattering
amplitude possesses the reality property, i.e., for $t$ real it is real between
the cuts and takes complex conjugate values above and below the cuts.

In the most favourable cases, dispersion relations have been established for
$-T < t \leq 0$  $T > 0$. A list of these cases have been given in 1958 by
Goldberger \cite{hh} and has not been enlarged since then. It is given in the
Table.

In the general case, even if dispersion relations are not proved, the crossing
property of Bros, Epstein and Glaser states that the scattering amplitude is
analytic in a twice cut plane, minus a finite region, for \underline{any
negative} $t$
\cite{jj}. So it is possible to continue the amplitude directly from
$A+B\rightarrow A+B$ to the complex conjugate of $A+\bar B\rightarrow A+\bar
B$. By a more subtle argument, using a path with fixed $u$ and fixed $s$ it is
possible to continue directly from
$A+ B\rightarrow A+ B$ to $A+\bar B\rightarrow A+\bar B$

At this point, we see already that one cannot dissociate analyticity, i.e.,
dispersion relations, and unitarity, since the discontinuity in the dispersion
relations is given by the absorptive part. In the simple case of $t = 0$, the
absorptive part is given by the total cross-section and the forward amplitude 
is given, as we said already for the case of Compton Scattering, by an
integral over physical quantities.

\begin{table}
\begin{center}
DISPERSION RELATIONS\\
\vspace{0.5cm}
a)  Proved relations \\ \vspace{0.3cm}
\begin{tabular}{|l|l||l|} \hline
Process & Limitation in invariant momentum  &
Continuation of absorptive \\
 & 
transfer &
 part into the unphysical \\
$k + p \rightarrow k^\prime + p^\prime$&&  region by convergent partial\\  && 
wave expansion  \\
\hline &&\\
$\pi + N \rightarrow \pi + N$ & $T{\rm max} = {8m_\pi^2\over 3}$~~
${2m_p+ m_\pi\over 2m_p-m_\pi} $& $ 0 \leq T < T ~{\rm max}$ \\ 
&& \\\hline
&&\\
$\pi + \pi \rightarrow \pi + \pi$ & $T~{\rm max} = 7 m^2_\pi$ (now 28$m^2_\pi$)
&
$ 0 
\leq T < T~{\rm max} $\\
&&\\ \hline
&& \\
$\gamma + N \rightarrow \gamma + N^{(*)}$ & 
$T~{\rm max} = \mu^2 \left\{{(2m_p + m_\pi)^2\over 4(m_p + m_\pi)^2}\right.
+
\left.{2m_p + m_\pi\over m_p}\right\}$ & 
$0 \leq T < T~{\rm max}$ \\ 
&& \\ \hline
&& \\
$\gamma + N \rightarrow \pi + N^{(*)}$ & $T~{\rm max} = F(0)^{(**)} \sim 3
m_\pi^2$ & $T th \leq T < T~{\rm max}$ \\
&& \\
$e + N \rightarrow e + \pi + N^{(*)}$ & $T~{\rm max} = F^{(**)}(\gamma);~~
\gamma \equiv k^2_0 - k^2$ & $T th = {m_p \over m_p + m_\pi}\times{m_\pi -
\gamma
\over 4}$ \\
&& \\
& $F(-9 m_\pi^2) \sim 6 m_\pi^2$ & \\
&& \\ \hline
\end{tabular} \\ \vspace{0.5cm}
b) Some unproved relations \\
\vspace{0.3cm}
\begin{tabular}{|l|l||l|} \hline
&& \\
\phantom{$e + N \rightarrow e + \pi + N^{(*)}$}& Mass restrictions appearing in
proof  &  Perturbation theory~~~~~~~~\\
&based upon causality and spectrum; &
(every finite order)
\\ 
& $T = 0$ &  \\
&& \\ \hline
&&\\
$N+N\rightarrow N+N$ & $m_\pi > (\sqrt{2}-1) m_p$ & proved for $T <
{m_\pi^2\over 4}$ \\
&& \\ \hline
&& \\
$K + N \rightarrow K + N$ & complicated; not fulfilled by narrow  & \\
&margin& \\
&&\\ \hline
&& \\
$\pi + D \rightarrow \pi + D$ & $\epsilon > {m_\pi \over 3}~;~~ m_D = 2m_p -
\epsilon$ & \\
&& \\ \hline
\end{tabular}   

\end{center}
\end{table}

It was recognized very early that the combination of analyticity and unitarity
might lead to very interesting consequences and might give some hope to fulfill
at least partially the $S$ matrix Heisenberg program. This was very clearly
stated already in 1956 by Murray Gell-Mann \cite{kk} at the Rochester
conference. Later this idea was taken over by many people, in particular by
Geff Chew. To make this program as successful as possible it seemed necessary
to have an analyticity domain
as large as possible. Dispersion relations are fixed $t$ analyticity
properties, in the other variable $s$, or $u$ as one likes.

Another property derived from local field theory was the existence of the
\underline{Lehmann ellipse} \cite{lll}, which states that for fixed $s$,
physical, the scattering amplitude is analytic in $\cos\theta$ in an ellipse
with foci at $\cos\theta = \pm 1$. $\cos \theta = 1$ corresponds to $t = 0$ the
ellipse therefore contains a circle 
\beq
\vert t \vert < T_1(s)
\label{ten}
\eeq
In the Lehmann derivation $T_1(s)\rightarrow 0$ for $s\rightarrow (M_A+M_B)^2$
and $s\rightarrow \infty$. 

The absorptive part is analytic in the larger ellipse, the ``large" Lehmann
ellipse, containing the circle 
\beq
\vert t \vert < T_2(s)
\label{eleven}
\eeq
with $T_2(s)\rightarrow c > 0$ for $s \rightarrow (M_A+M_B)^2$, $T_2(s)
\rightarrow 0$ for $s\rightarrow\infty$.

It was thought by Mandelstam that these two analyticity properties, dispersion
relations and Lehmann ellipses, were insufficient to carry very far the
analyticity-unitarity program. he proposed the Mandelstam representation
\cite{mm} which can be written schematically as

\vfill\eject

\bea
F = &&{1\over \pi^2} \int {\rho (s^\prime, t^\prime)ds^\prime dt^\prime \over
(s^\prime -s)~(t^\prime -t)}\nonumber \\
&&+ {\rm circular~permutations~in}~ s, t, u \nonumber \\
&&+ {\rm one~dimensional~dispersion~integrals} \nonumber \\
&&+ {\rm subtractions}
\label{twelve}
\eea

This representation is nice. It gives back the ordinary dispersion relations
and the Lehmann ellipse when one variable is fixed, but it was never proved nor
disproved for all mass cases, even in perturbation theory,. One contributor,
Jean Lascoux, refused to co-sign a ``proof", which, in the end, turned out to be
imperfect.

One very impressive consequence of Mandelstam representation was the proof, by
Marcel Froissart, that the total cross-section cannot increase faster than (log
$s)^2$, the so-called ``Froissart Bound" \cite{nn}.

My own way to obtain the Froissart bound \cite{oo} was to use the fact that the
Mandelstam representation implies the existence of an ellipse of analyticity in
$\cos\theta$ qualitatively \underline{larger} than the Lehmann ellipse, i.e.,
such that it contains a circle $\vert t\vert < R$, $R$ fixed, independent of
the energy. This has a consequence that ${\rm Im}~ f_\ell(s)$ decreases with
$\ell$ at a certain exponential rate because of the convergence of the Legendre
polynomial expansion and of the polynomial boundedness, but on the other hand
the ${\rm Im}~ f_\ell(s)$'s are bounded by unity because of unitarity [Eq.
(\ref{seven})]. taking the best bound for each $\ell$ gives the Froissart bound.

To prove the Froissart bound without using the Mandelstam representation one
must find a way to enlarge the ``small" and the ``large" Lehmann ellipses. In
the autumn of 1965, I had very stimulating discussions with Harry Lehmann
at the ``Institut des Hautes Etudes Scientifiques" about an attempt made in
this direction by Nakanishi in which he combined in a not very consistent way
positivity and some analyticity properties derived from perturbation theory. He
was using a domain shrinking to zero when the energy became physical and this
lead nowhere. Finally, in December 1965 \cite{pp}, I found the way out. The
positivity of ${\rm Im}~ f_\ell$ implies, by using expansion (\ref{five}),
\beq
\left\vert \left({d\over dt}\right)^n A_S(s,t)\right\vert
_{-4k^2 \leq t \leq 0}
 \leq
\left\vert\left({d\over dt}\right)^n A_S (s,t) \right\vert_{t=0}
\label{thirteen}
\eeq
To calculate 
$$F(s,t) = {1\over\pi}~~\int_{s_0} {A_s(s^\prime t) ds^\prime \over s^\prime -s}
$$
(forget the left-hand cut and subtractions!), for $s$ real $< s_0$ one can
expand $F(s,t)$ around $t = 0$. From the property (\ref{thirteen}) one can
prove that the successive derivatives can be obtained by differentiating
\underline{under the integral}. When one resums the series one discovers that
this can be done not only for $s$ real $<s_0$, but for any $s$ and that the
expansion has a domain of convergence in $t$ \underline {independent of $s$}.
This means that the large Lehmann ellipse \underline{must} contain a circle
$\vert t \vert < R$. This is exactly what is needed to get the Froissart bound.
In fact, in favourable cases, $R = 4m^2_\pi$, $m_\pi$ being the pion mass. A
recipe to get a lower bound for $R$ was found by Sommer \cite{qq}
\beq
R \leq {\rm sup}_{s_0 < s <\infty} T_1(s)
\label{forteen}
\eeq 
It was already
known that for
$\vert t
\vert < 4m^2_\pi$ the number of subtractions in the dispersion relations was at
most
\underline{two}
\cite{rr}, and is lead to the more accurate bound \cite{ss}
\beq
\sigma_T < {\pi\over m^2_\pi}~~(\log s)^2
\label{fifteen}
\eeq

Notice that this is \underline{only} a \underline{bound}, not an asymptotic
estimate.

In spite of many efforts the Froissart bound was never qualitatively improved,
and it was shown by Kupsch \cite{tt} that if one uses only ${\rm Im}~ f_\ell
\geq
\vert f_\ell\vert^2$ and full crossing symmetry one cannot do better than
Froissart. 

Before 1972, rising cross-sections were a pure curiosity. Almost everybody
believed that the proton-proton cross-section was approaching 40 millibarns at
infinite energy. Only Cheng and Wu \cite{uu} had a QED inspired model in which
cross-sections were rising and behaving like $(\log s)^2$ at extremely high
energy. Yet, Khuri and Kinoshita \cite{vv} took seriously very early the
possibility that cross-sections rise and proved, in particular, that if the
scattering amplitude is dominantly crossing even, and if $\sigma_t \sim (\log
s)^2$ then
$$
\rho = {Re F\over Im F} \sim {\pi\over \log s}~,
$$
where $Re F$ and $Im F$ are the real and imaginary part of the forward
scattering amplitude.

In 1972, it was discovered at the  ISR, at CERN, that the $p-p$ cross-section
was rising by 3 millibarns from 30 GeV c.m. energy to 60 GeV c.m. energy
\cite{ww}. I suggested to the experimentalists that they should measure $\rho$
and test the Khuri-Kinoshita predictions. They did it \cite{yy} and this kind
of combined measurements of $\sigma_T$ and $Re F$ are still going on. In
$\sigma_T$ we have now more than a 50 \% increase with respect to low energy
values. For an up to date review I refer to the article of Matthiae
\cite{zut}. it is my \underline{strong conviction} that this activity should be
continued with the future LHC. A breakdown of dispersion relation might be a
sign of new physics due to the presence of extra compact dimensions of space
according to N.N. Khuri \cite{zz}. Future experiments, especially for $\rho$,
will be difficult because of the necessity to go to very small angles, but not
impossible \cite{aai}.

Before leaving the domain of high-energy scattering I would like to indicate
the new version of the Pomeranchuk theorem. When it was believed that
cross-sections were approaching finite limits, the Pomeranchuk theorem
\cite{bbi} stated that, under a certain assumption on the real part
$$
\sigma_T(AB) - \sigma_T(A\bar B) \rightarrow 0
$$

If cross-sections are rising to infinity, one can actually prove, according to
Eden \cite{cci} and Kinoshita \cite{ddi} that
$$
\sigma_T(AB) / \sigma_T(A\bar B) \rightarrow 1~.
$$

Now I would like to turn to another aspect of analyticity-unitarity. A
consequence of the enlargment of the Lehmann ellipse is that, in the special
case of $\pi\pi\rightarrow\pi\pi$ scattering, one can, by using crossing
symmetry, obtain a very large analyticity domain \cite{eei}, but one can prove
that the domain is smaller than the Mandelstam domain \cite{ffi}. By playing
with crossing symmetry and unitarity in a clever way (with years enormous
progress has been made according to the Figure), one gets a bound on the
scattering amplitude at the ``symmetry point" which is \cite{ggi}
$$
\left\vert F (s = t = u = 4m^2_\pi/3\right\vert < 4~,
$$

\begin{figure}[htb]
\hglue 5cm
 \epsfig{figure=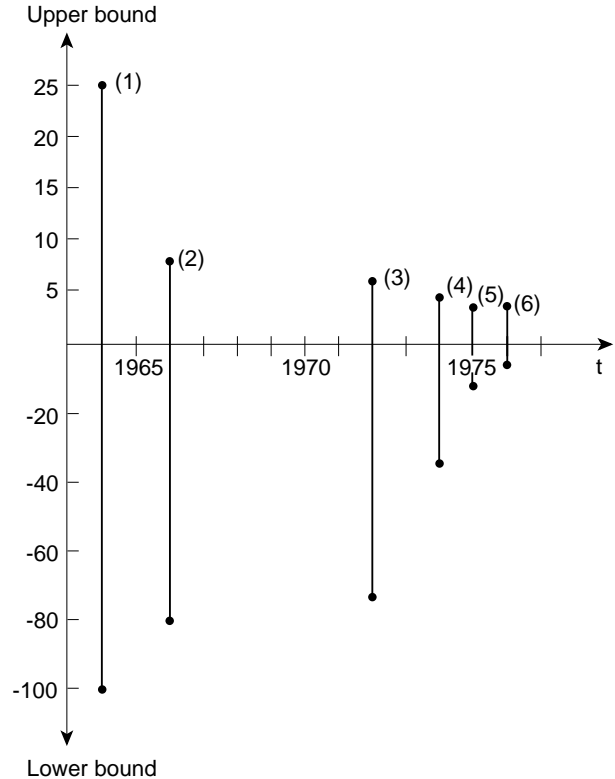,width=8cm}
 \caption[]{Bounds on the scattering amplitude at the symmetry point $s = t = u
= 4/3m^2_\pi$ as a function of time. Normalization: $F(s=4m^2_\pi, 0, 0)$ =
scattering length.}
\end{figure}

where $F$ is normalized in such a way that $F (s=u, t = 0, u=0)$ is the
$\pi_0\pi_0$ scattering length, $a_{00}$.
One can also obtain a lower bound on the scattering length, the bound value
being \cite{hhi}
$$
a_{00} > -1.75~~(m_\pi)^{-1}~,
$$
a number which is off the model predictions only by a factor 10.

Though these latter results may seem ``useless", they are remarkable, since they
prove that the combination of analyticity and unitarity have a dynamical
content.

\vfill\eject

\end{document}